\documentclass{PoS}

\usepackage{amsfonts}
\usepackage{amsmath}
\usepackage{subfigure} 

\graphicspath{{./figs/}}

\newcommand{\Tr}{\textrm{Tr}}

\title{
Non-Perturbative Renormalization for Staggered Fermions
(Self-energy Analysis)
}

\ShortTitle{ Staggered NPR }

\author{\speaker{Jangho Kim}, Boram Yoon,
and Weonjong Lee \\
Lattice Gauge Theory Research Center, CTP, and FPRD, \\
Department of Physics and Astronomy, \\
Seoul National University, Seoul, 151-747, South Korea \\
E-mail: \email{wlee@snu.ac.kr}}

\author{SWME Collaboration}

\abstract{
We present preliminary results of data analysis for the
non-perturbative renormalization (NPR) on the self-energy
of the quark propagators calculated using HYP improved
staggered fermions on the MILC asqtad lattices.
We use the momentum source to generate the quark propagators.
In principle, using the vector projection operator of
$(\overline{\overline{ \gamma_\mu \otimes 1}})$ and the scalar
projection operator $(\overline{\overline{ 1 \otimes 1}})$ ,
we should be able to obtain the wave function renormalization factor
$Z_q'$ and the mass renormalization factor $Z_q \cdot Z_m$.
Using the MILC coarse lattice, we obtain a preliminary but reasonable
estimate of $Z_q'$ and $Z_q\cdot Z_m$ from the data analysis on the
self-energy.  
}

\FullConference{
The 30th International Symposium on Lattice Field Theory - Lattice 2012\\
June 24-29, 2012\\
Cairns Convention Centre, Cairns, Australia
}

\begin{document}

\section{Introduction \label{sec:intr}} 
We can obtain the wave function renormalization factor $Z_q'$ defined
in the RI${}^\prime$-MOM scheme and the mass renormalization factor
$Z_q \cdot Z_m$ defined in the RI-MOM scheme from the staggered quark
propagator using non-perturbative renormalization (NPR) method.
We generate the staggered quark propagators using the momentum source
in the Landau gauge on the MILC coarse lattices \cite{lytle,lytle2}.
Here, we present results of the data analysis after the projection.

\section{Mass Renormalization}
Let us consider a staggered fermion propagator.
\begin{eqnarray}
S^{f}_{cc'}(x_{1},x_{2}) &&\equiv \langle \chi^{f}_{c}(x_{1}) 
\overline{\chi}^{f}_{c'}(x_{2}) \rangle \,,
\end{eqnarray}
where $f$ is a flavor index, $c$, $c'$ are color indices.
Here, $x_{1}$ and $x_{2}$ represent the position coordinates on the
lattice with $x_{1}, x_{2}\in \mathbb{Z}^{4}$.
The lattice spacing is $a$.

In the normal Brillouin zone, we use $p,q$ as the momentum, and, in
the reduced Brillouin zone, we use $\tilde{p}$, $\tilde{q}$ as the
momentum as follows,
\begin{eqnarray}
&&p,q \in (-\frac{\pi}{a},\frac{\pi}{a}]^{4}\,, \qquad
\tilde{p},\tilde{q} \in (-\frac{\pi}{2a},\frac{\pi}{2a}]^{4}\,,\qquad
q = \tilde{q}+\pi_{A}\,, \qquad
p = \tilde{p}+\pi_{B}\,,
\end{eqnarray}
where $A$, $B$ are hypercubic vectors whose element is $0$ or
$1$, and $\displaystyle{\pi_{A} \equiv \frac{\pi}{a}A}$.

The staggered quark propagator is defined as
\begin{eqnarray}                                                               
\hat{S}^{f}_{cc'}(\tilde{p}+\pi_{A},\tilde{q}+\pi_{B})
&\equiv&
\langle \widetilde{\chi}^{f}_{c}(\tilde{p}+\pi_{A})\, 
\widetilde{\overline{\chi}}^{f}_{c'}
(\tilde{q}+\pi_{B}) \rangle \,.
\end{eqnarray}
Using the Fourier analysis, one can show the following relationship.
\begin{eqnarray}
\hat{S}^{f}_{cc'}(\tilde{p}+\pi_{A},\tilde{q}+\pi_{B}) 
=\widetilde{\delta}^{(4)}(\tilde{p}-\tilde{q})
[\widetilde{S}(\tilde{p})]^{f}_{AB;cc'} 
\end{eqnarray}
where 
\begin{eqnarray}
\widetilde{\delta}^{(4)}(\tilde{p}) 
&\equiv& 
(2a)^{4}\sum_{y \in \mathbb{W}^{4} }e^{i\tilde{p}y}
\end{eqnarray}
Here, $y$ represents a position coordinate of a hypercube whose
lattice spacing is $2a$, $y\in \mathbb{W}^{4}$, and $\mathbb{W}$
denotes one-dimensional lattice whose spacing is $2a$.
By setting $\tilde{p}=\tilde{q}$, the quark propagator becomes
\begin{eqnarray}
\hat{S}^{f}_{cc'}(\tilde{p}+\pi_{A},\tilde{p}+\pi_{B}) 
&&= \widetilde{\delta}^{(4)}(0) [\widetilde{S}(\tilde{p})]^{f}_{AB;cc'}
= V [\widetilde{S}(\tilde{p})]^{f}_{AB;cc'}\,,
\end{eqnarray}
where $V$ is lattice volume factor.
\begin{eqnarray}
V \equiv (2a)^{4}\sum_{y \in \mathbb{W}^{4} } 1
= L^3 \times T
\end{eqnarray}
where $L$ ($T$) is lattice size in the spacial (time) direction.

To obtain the propagator, we have to solve the staggered 
Dirac equation.
\begin{eqnarray}
&&(\displaystyle{\not}{D}_{s}+m^{f})_{i}\psi^{f}_{i}=h \nonumber\\
&&\psi^{f}_{i}=\frac{1}{(\displaystyle{\not}{D}_{s}+m^{f})_{i}}h\,,
\end{eqnarray}
where $i$ is a gauge configuration index,
$(\displaystyle{\not}{D}_{s}+m^{f})_{i}$ is Dirac operator for
staggered fermion.
$h$ is a source vector, and $\psi^{f}_{i}$ is a solution vector of the
Dirac equation for a specific gauge configuration $i$.
\begin{eqnarray}
S^{f}_{i}(x_{1},x_{2}) = \frac{1}{(\displaystyle{\not}{D}_{s}+m^{f})_{i}}
\end{eqnarray}
where $S^{f}_{i}(x_{1},x_{2})$ is a propagator for a specific gauge 
configuration $i$.
A thermalized quark propagator is defined as 
\begin{eqnarray}
&&S^{f}_{cc'}(x_{1},x_{2})=
\frac{1}{N} \sum^{N}_{i} S^{f}_{i;cc'} (x_{1},x_{2})\,,
\end{eqnarray}
where $N$ is the number of gauge configurations.

\begin{figure}[!h]
\centering
\includegraphics[width=34pc]{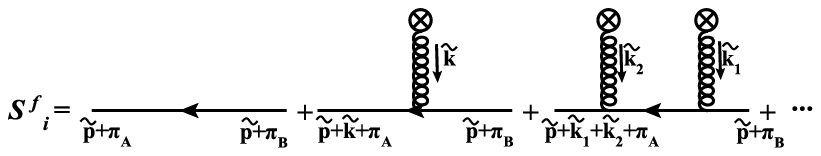}
\includegraphics[width=34pc]{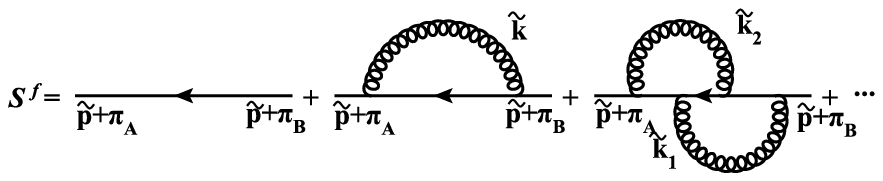}
\caption{$S^{f}_{i}$ has temporal external gluons, while 
	all the gluons are contracted in $S^{f}$.
}
\label{Si_stag}
\end{figure}

$S^{f}$ and $S^{f}_{i}$ are quite different.
$S^{f}_{i}$ do not conserve the momentum, because there are external
gluons which live for a short time in a Monte Carlo evolution time.
But $S^{f}$ conserve the momentum in the reduced Brillouin zone.
The difference is shown in Figure~\ref{Si_stag}.

The solution vector $\psi^{f}_{i}$ is
\begin{eqnarray}
&&\psi^{f}_{i;a}(x_{1})=a^{4}\sum_{x_{2}\in \mathbb{Z}^{4} }
S^{f}_{i;ab}(x_{1},x_{2})h_{b}(x_{2})
\end{eqnarray}
We set the source vector 
$h_{b}(x_{2},\tilde{p}+\pi_{B})=e^{-i(\tilde{p}+\pi_{B})x_{2}}
\delta_{bc}$.
\begin{eqnarray}
&&\psi^{f}_{i,ac}(x_{1},\tilde{p}+\pi_{B})
=a^{4}\sum_{x_{2} \in \mathbb{Z}^{4}}
S^{f}_{i;ab}(x_{1},x_{2})e^{-i(\tilde{p}+\pi_{B})x_{2}}
\delta_{bc}\nonumber\\
\end{eqnarray}
After we calculate $\psi^{f}_{i;ac}(x_{1},\tilde{p}+\pi_{B})$ 
using conjugate gradient method for each
$c$, we can obtain the full matrix of 
$\psi^{f}_{i;ab}(x_{1},\tilde{p}+\pi_{B})$:
\begin{eqnarray}
\hat{S}^{f}_{ab}(\tilde{p}+\pi_{A},\tilde{p}+\pi_{B})
&&=\frac{a^{4}}{N}\sum^{N}_{i}\sum_{x_{1} \in \mathbb{Z}^{4} }
e^{i(\tilde{p}+\pi_{A})x_{1} }\psi^{f}_{i;ab}(x_{1},
\tilde{p}+\pi_{B})
=V[\widetilde{S}(\tilde{p})]^{f}_{AB;ab}
\end{eqnarray}
The inverse bare propagator can be expressed as follows, 
\begin{eqnarray}
&&[\widetilde{S}^{f}(\tilde{p})]^{-1}_{AB;cc'}
=[{(1+\Sigma_{S})}m_{0}^{f}\overline{\overline{(1 \otimes 1)}}_{AB} 
+ {(1+\Sigma_{V})}\sum_{\mu}\frac{i}{a}\sin{(\tilde{p}_{\mu}a)}
\overline{\overline{(\gamma_{\mu} \otimes 1)}}_{AB} \nonumber\\
&&\phantom{11111111111}+{\Sigma_{T}} m_{0}^{f} 
\sum_{\mu \ne \nu} \sin{(\tilde{p}_{\mu}a)} 
(\sin{(\tilde{p}_{\nu}a}))^{3}
\overline{\overline{(\gamma_{\mu\nu} \otimes 1)}}_{AB} \nonumber\\
&&\phantom{11111111111}+{\Sigma_{A}}
\sum_{\mu \ne \nu \ne \rho} 
\frac{i}{a}\sin{(\tilde{p}_{\mu}a)} (\sin{(\tilde{p}_{\nu}a)})^{3} 
(\sin{(\tilde{p}_{\rho}a)})^{5}
\overline{\overline{(\gamma_{\mu\nu\rho} \otimes 1)}}_{AB} \nonumber\\
&&+{\Sigma_{P}}m_{0}^{f}\sum_{\mu \ne \nu \ne \rho \ne \sigma}
\sin{(\tilde{p}_{\mu}a)}
(\sin{(\tilde{p}_{\nu}a)})^{3}
(\sin{(\tilde{p}_{\rho}a)})^{5}
(\sin{(\tilde{p}_{\sigma}a)})^{7}
\overline{\overline{(\gamma_{\mu\nu\rho\sigma} \otimes 1)
}}_{AB} ]_{cc'}\nonumber
\end{eqnarray}
which is derived from the lattice symmetry \cite{sharpe}.
The definition of $\overline{\overline{(\gamma_{S} 
\otimes \xi_{F})}}_{AB}$ is given as
\begin{eqnarray}
&&\overline{(\gamma_{S} \otimes \xi_{F})}_{AB} \equiv 
\frac{1}{4}Tr[\gamma^{\dagger}_{A}\gamma_{S}\gamma_{B}
\gamma^{\dagger}_{F}]\nonumber\\	
&&\overline{\overline{(\gamma_{S} \otimes \xi_{F})}}_{AB} \equiv 
\frac{1}{16}\sum_{C,D}(-1)^{A \cdot C}
\overline{(\gamma_{S} \otimes  \xi_{F})}_{CD} 
(-1)^{D \cdot B} 
\end{eqnarray}
The renormalization of propagator is
$\widetilde{S}^{f}_{R}(p) = Z_{q}\widetilde{S}^{f}_{0}(p)$,
and the mass renormalization is defined by $m_{R}=Z_{m}m_{0}$.
Here, $Z_{q}$ is wave function renormalization factor for quark field,
$Z_{m}$ is mass renormalization factor, $m_{R}$ is renormalized quark
mass, $m_{0}$ is a bare quark mass and the subscript $R$ denotes a
renormalized quantity, the subscript $0$ denotes a bare quantity.
Unless specified, we use the convention of $m = m_0$ in this paper.

The RI${}^\prime$-MOM scheme prescription is 
\begin{align}
  & Z'_q = -i \frac{1}{48} \sum_\mu
  \frac{\hat{p}_\mu}{\hat{p}^2} 
  \Tr[(\overline{ \overline{\gamma_\mu \otimes 1}}) S^{-1}(\tilde{p})]
  \label{Eq:Z_q}
  \\ 
  & Z'_q \left[ Z_m m  + C_1 \frac{\langle\bar\chi\chi\rangle}{\hat{p}^2}
    \right]
  = \frac{1}{48}
  \Tr[(\overline{ \overline{1 \otimes 1}}) S^{-1}(\tilde{p})] \,,
\end{align}
where $\hat{p}_\mu \equiv \sin(a\tilde{p}_\mu)$ and 
$\hat{p}^2 \equiv \sum_\mu \hat{p}_\mu^2$.
So the renormalized propagator can be rewritten as
\begin{eqnarray}
&&\widetilde{S}^{f}_{R}(p) = 
\frac{Z_{q}}{(1+\Sigma_{V})}
\left ({\displaystyle{\sum_{\mu}}
\displaystyle{\frac{i}{a}}\sin{(\tilde{p}_{\mu}a)}
\overline{\overline{(\gamma_{\mu} \otimes 1)}}_{AB}
+\displaystyle{\frac{(1+\Sigma_{S})}{(1+\Sigma_{V})}}
\frac{m_{R}}{Z_{m}}\overline{\overline{(1 \otimes 1)}}_{AB}}
+\cdots \right)^{-1}\nonumber\\
\end{eqnarray}
Thus, we can write the $Z_q$ and $Z_m$ as follows,
\begin{eqnarray}
Z_{q} = (1+\Sigma_{V})\,,\qquad
Z_{m} = \frac{(1+\Sigma_{S})}{(1+\Sigma_{V})}.
\end{eqnarray}
\section{Results}
\label{sec:result}
\begin{figure}[!t]
\center
\subfigure[$\Delta r_V$]{
\includegraphics[width=0.48\textwidth]{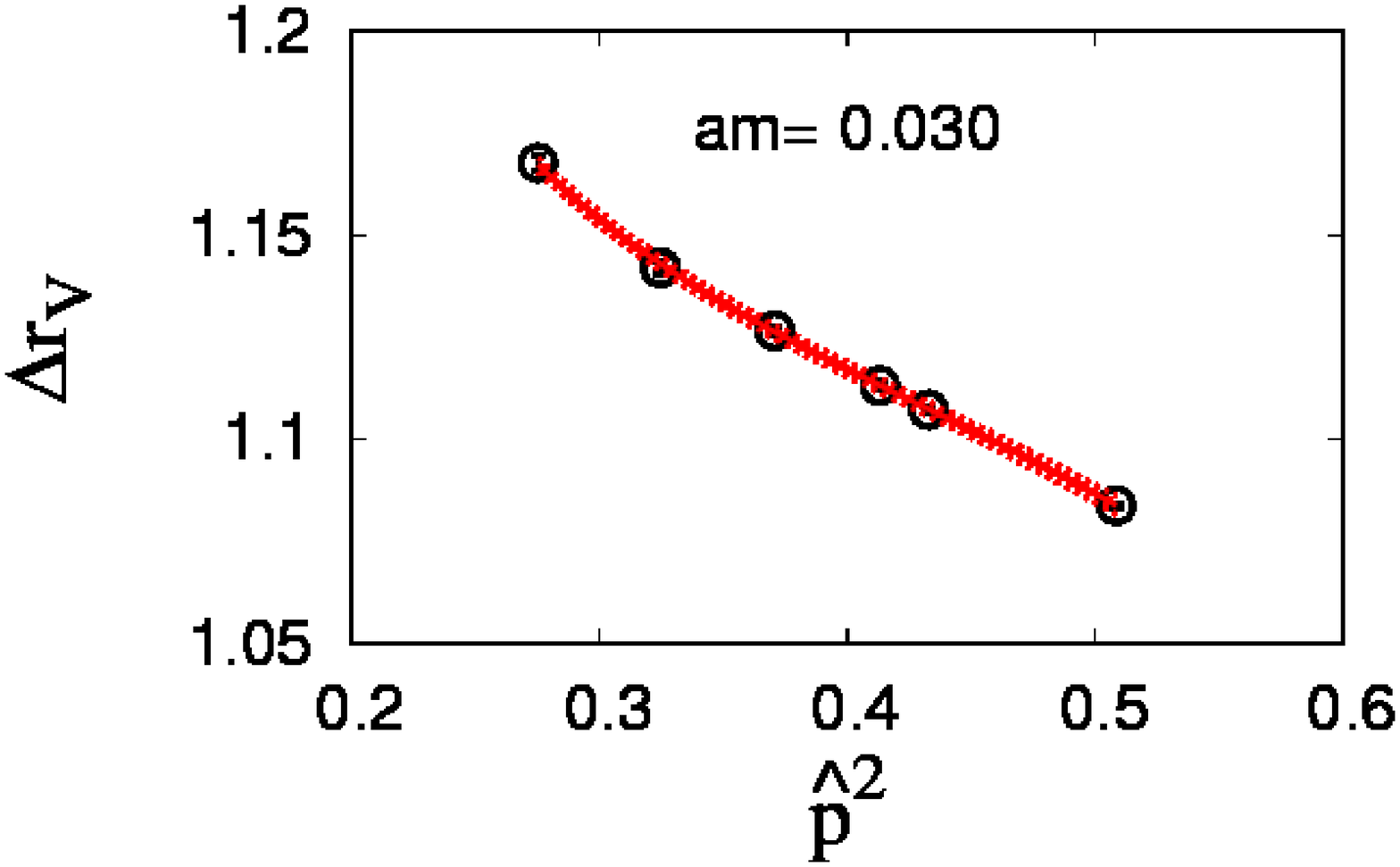}
\label{fig:VxSvsMOM}}
\subfigure[$y_V$]{
\includegraphics[width=0.48\textwidth]{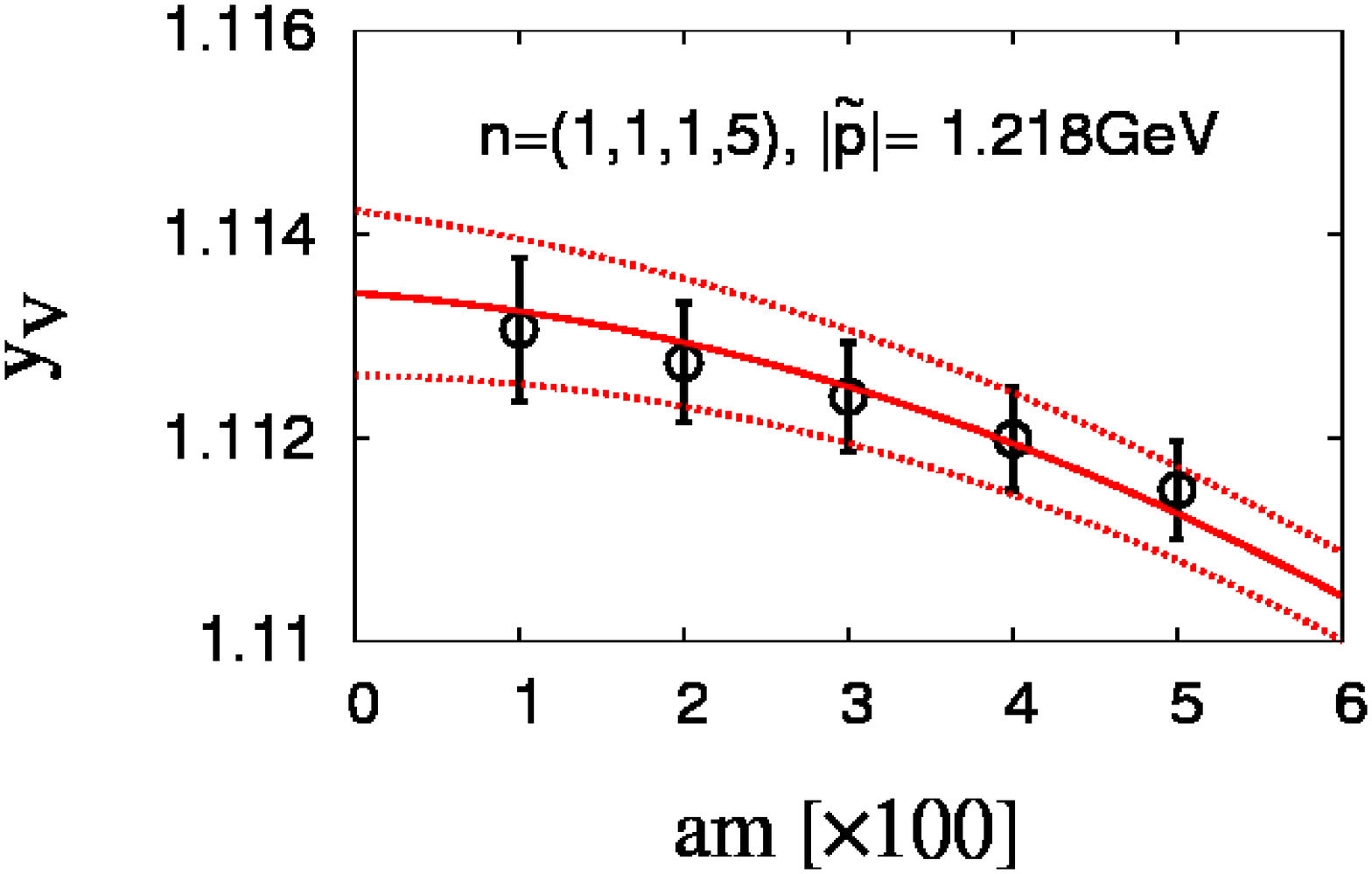}
\label{fig:VxSvsMASS}}
\caption{
$\Delta r_V$ vs. $\hat{p}^2$ (left) and  
$y_V$ vs. $am$ (right). 
}
\end{figure}

%
%
%
%
%
We generate staggered fermion propagators 
for 5 quark masses and 6 external momenta with $0.5 < |a\tilde{p}| < 0.75$. 
$am = 0.01, 0.02, 0.03, 0.04, 0.05$. 
%
%
%
The fitting function suggested in
Ref.~\cite{wlee,politzer,blum,RBC-UKQCD} is
\begin{align}
y_i & = \dfrac{1}{N} \Tr [S^{-1}(\tilde{p}) \, \mathbb{P}_i]
\\
f_q(m,a,\hat{p}) &=
c_1 \left(1+\Gamma_1 [\log(\hat{p}^2) + 2\dfrac{(am)^2}{\hat{p}^2}]\right)
+ c_2 \log(\hat{p}^2) + c_3 [\log(\hat{p}^2)]^2
+ c_4 \dfrac{(am)^2}{\hat{p}^2} \\
&+ c_5 \dfrac{(am)^2}{\hat{p}^2}\log(\hat{p}^2) 
+ c_6 (am) + c_7 \hat{p}^2 + c_8 (\hat{p}^2)^2 + c_9 \hat{p}^4 
\\
f_m(m,a,\hat{p}) &= \dfrac{d_1}{\hat{p}^2} 
+ (am)\bigg( d_2(1+\Gamma_2[\log(\hat{p}^2) + \dfrac{(am)^2}{\hat{p}^2}
+ \dfrac{(am)^2}{\hat{p}^2}\log(1+\dfrac{\hat{p}^2}{(am)^2}) ] ) \\ 
& + d_3 \log(\hat{p}^2) + d_4 [\log(\hat{p}^2)]^2 
+ d_5 \dfrac{(am)^2}{\hat{p}^2}
+ d_6 \dfrac{(am)^2}{\hat{p}^2}\log(\hat{p}^2)
\\
& + d_7 \hat{p}^2 + d_8 (\hat{p}^2)^2 + d_9 \hat{p}^4  \bigg)
\end{align}
where the anomalous dimension $\Gamma_i$ is
\begin{align}
\Gamma_1 & = -\dfrac{\alpha_s}{(4\pi)} \cdot \dfrac{4}{3} \,,
\qquad 
\Gamma_2 = -\dfrac{\alpha_s}{4\pi} \cdot \dfrac{16}{3} \,.
\end{align}
Here, $y_i$ represents a data point obtained by some projection
$\mathbb{P}_i$.
Let us consider a data analysis for $Z'_q$ with a vector projection:
$\mathbb{P}_V = \overline{\overline{(\gamma_\mu \otimes 1)}} 
\hat{p}/\hat{p}^2$.
We use the uncorrelated Bayesian method to fit the data to $f_q(X)$ by
imposing the following prior condition: $c_1 = 1 \pm 0.5 \alpha_s$,
$c_{2-5} = 0 \pm 2 \alpha_s^2$, and $c_{6-9} = 0 \pm 2$.
Here, $X$ represents $m,\hat{p},a$ collectively
Here, note that the prior information on $c_{1-5}$ comes from the
lattice perturbation theory \cite{wlee}.
In order to investigate the fitting quality, let us define $\Delta
r_V$ as $\Delta r_V \equiv y_V - c_9 \hat{p}^4$.
%
%
%
In Fig.~\ref{fig:VxSvsMOM} and \ref{fig:VxSvsMASS}, we present
$\Delta r_V$ and $y_V$, respectively.
The fitting results are given in Table~\ref{tab:z_q-fit-1}.
As you can see in the plots, the fitting quality is quite good.
\begin{table}[!b]
  \caption{ Fitting results for $Z_q'$.}
\label{tab:z_q-fit-1}
\center
\begin{tabular}{l  l  l  l  l}
\hline
\hline
$c_1$ & $c_2$ & $c_3$ & $c_4$ & $c_5$  \\
\hline
     0.931( 18)
  &  0.185( 47)
  &  0.161( 28)
  & -0.096( 18)
  &  0.263( 39)\\
\hline
\hline
$c_6$ & $c_7$ & $c_8$& $c_9$ & $\chi^2/\text{d.o.f}$  \\
\hline
     -0.012( 13)
  &   0.97( 17)
  &  -1.17( 19)
  &   0.373( 27)
  &   0.35( 11)\\
\hline
\hline
\end{tabular}
\end{table}

\begin{figure}[!t]
\center
\subfigure[$\Delta r_S$]{
\includegraphics[width=0.49\textwidth]{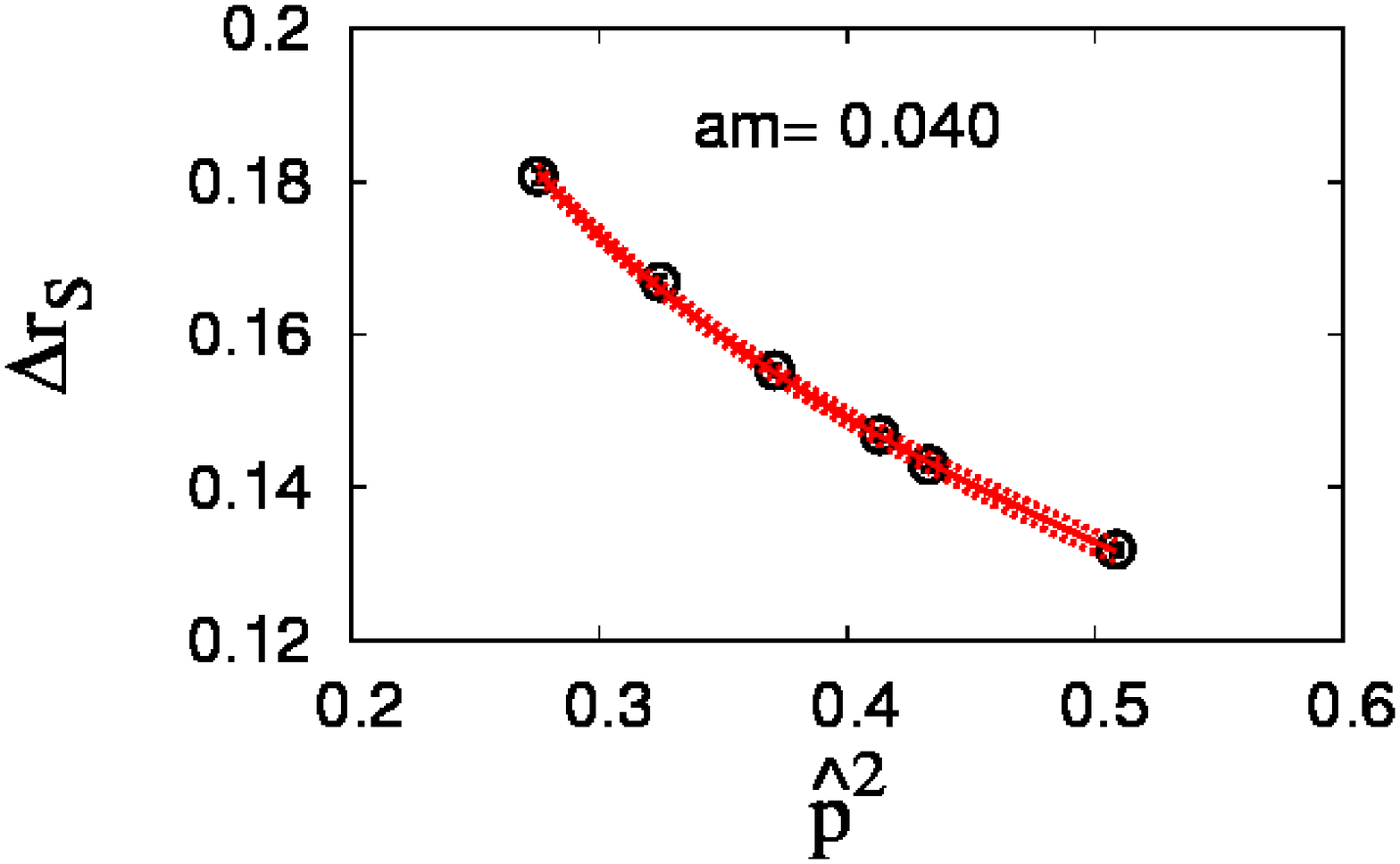}}
\subfigure[$y_S$]{
\includegraphics[width=0.49\textwidth]{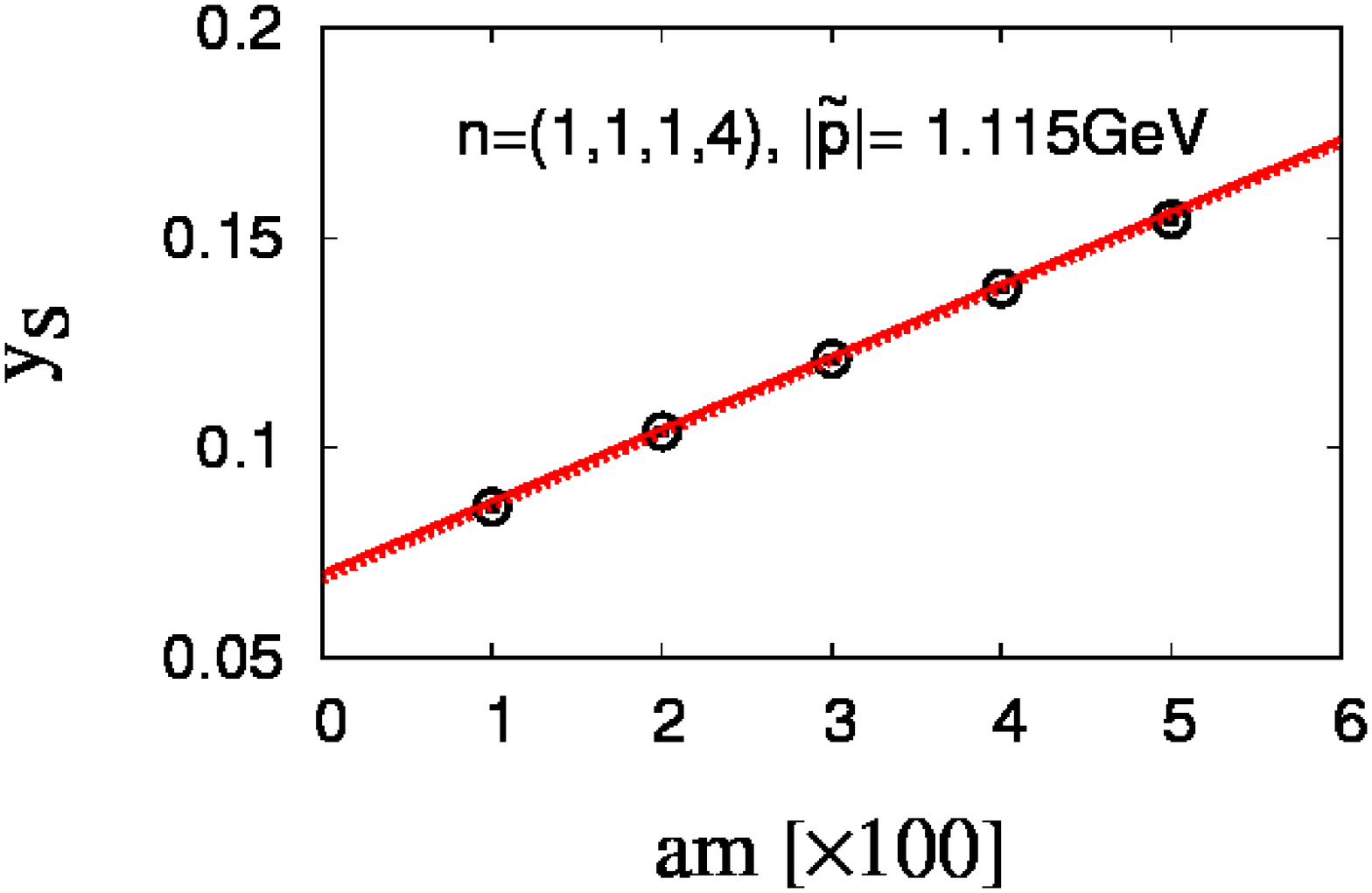}}
\caption{
    $\Delta r_S$ vs. $\hat{p}^2$ (left), and 
    $y_S$ vs. $am$ (right).
}
\label{fig:SxSvsMOM}
\end{figure}

Let us turn to the data analysis for the scalar projection:
$\mathbb{P}_S = (\overline{\overline{1 \otimes 1 }})$.
We use the Bayesian method to fit the data to $f_m(X)$.
The prior conditions are $d_2 = 1 \pm 0.5 \alpha_s$,
$d_{3-6} = 0 \pm 2 \alpha_s^2$, and $d_{7-9} = 0 \pm 2$.
Let us define
$\Delta r_S$ as $\Delta r_S \equiv y_S - d_9 (am) \hat{p}^4$.
%
%
In Fig.~\ref{fig:SxSvsMOM}, we show $\Delta r_S$ and $y_S$.
As one can see in the plots, the fitting quality is somewhat poor with
$\chi^2/\text{d.o.f} = 1.24(39)$ for the uncorrelated Bayesian
fitting.
The fitting results are summarized in Table \ref{tab:z_qz_m}.
\begin{table}[!b]
\caption{ Fitting results for $Z_q \cdot Z_m$.
}
\label{tab:z_qz_m}
\center
\begin{tabular}{l  l  l  l  l}
\hline
\hline
$c_1$ & $c_2$ & $c_3$ & $c_4$ & $c_5$  \\
\hline
    0.02992( 38)
  & 1.143( 11)
  &-0.205( 19)
  &-0.117( 20)
  &-0.0346( 82)       \\
\hline
\hline
$c_6$ & $c_7$ & $c_8$& $c_9$ & $\chi^2/\text{d.o.f}$ \\
\hline
    0.0414( 98)
  &2.358( 82)
  &-2.66( 16)
  &-2.55( 39)
  &1.24( 39)        \\
\hline
\hline
\end{tabular}
\end{table}
\begin{figure}[!htbp]
\center
\subfigure[$Z_q'$]{
\includegraphics[width=0.49\textwidth]{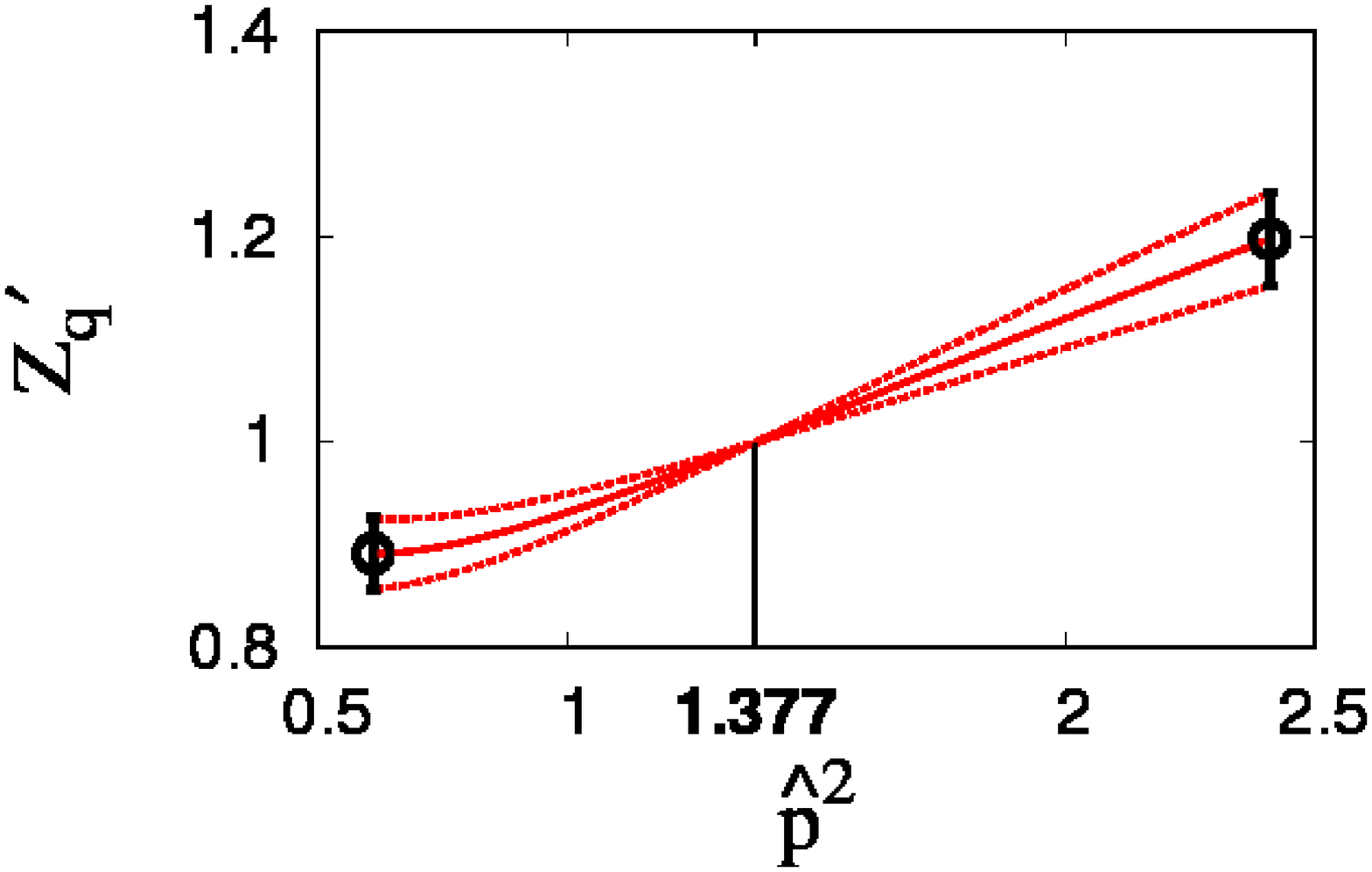}}
\subfigure[$Z_q \cdot Z_m$]{
\includegraphics[width=0.49\textwidth]{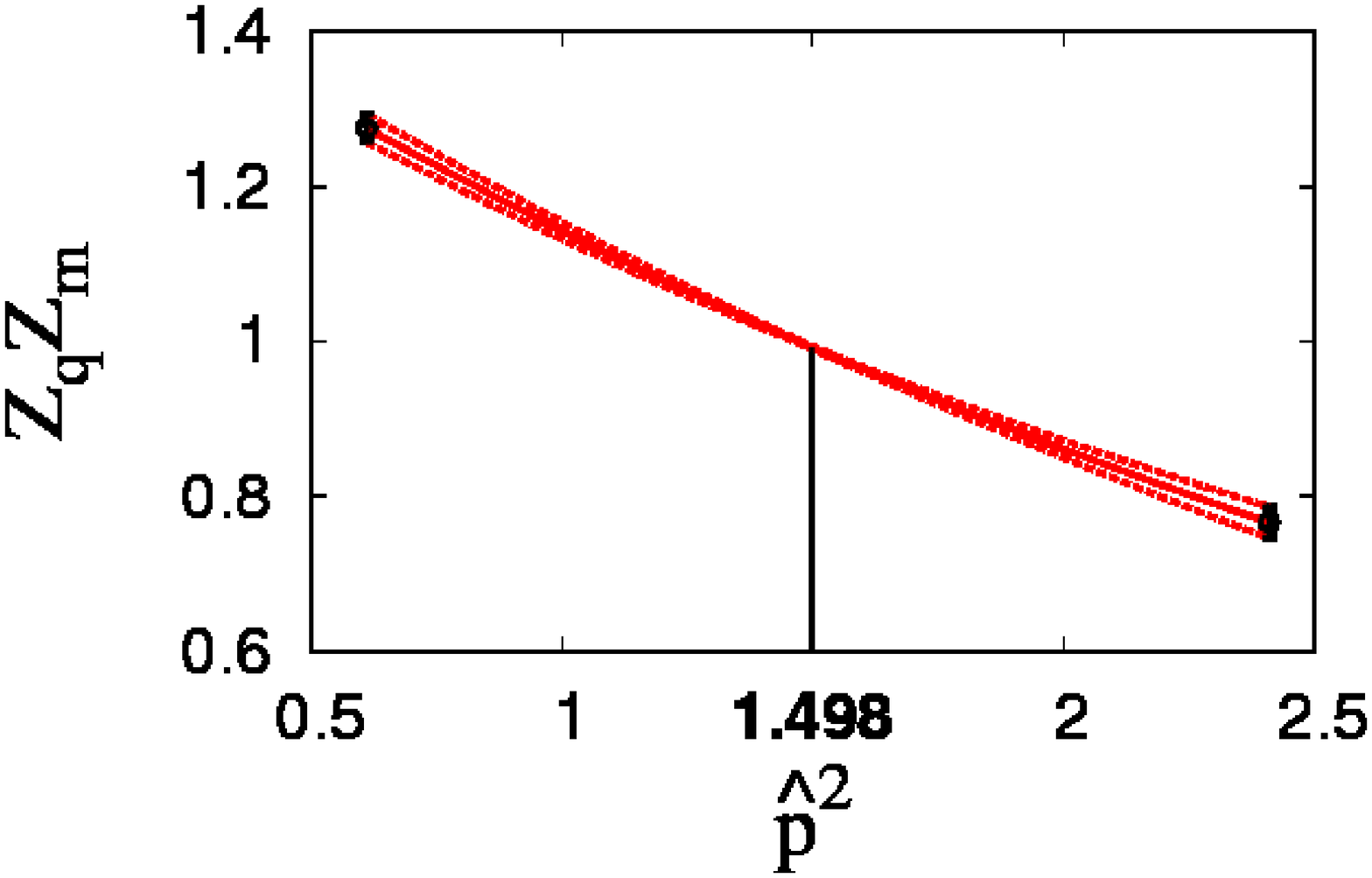}}
\caption{
  $Z_q'$ (left) and $Z_q \cdot Z_m$ (right) as a function
  of $\hat{p}^2$.
  The solid curve represents the central value and the dotted
  curves represent the statistical error.
}
\label{fig:z_qz_m}
\end{figure}
In Fig.~\ref{fig:z_qz_m}, we define the y-axis variables as
\begin{align}
Z_q'(\tilde{p}) & \equiv f_q(\,\tilde{p}; \,m=0, \,c_{7-9}=0)
\\
Z_q \cdot Z_m(\tilde{p}) & \equiv \frac{1}{am}
f_m(\,\tilde{p};\,m=0, \,d_1=0, \,d_{7-9}=0)
\end{align}
We estimate the statistical errors using the jackknife resampling
method.
As you can see in the plots, the minimum of statistical errors
is located at $|\tilde{p}| = 2.084 \text{GeV}$ for $Z_q'$ and
at $|\tilde{p}| = 2.190 \text{GeV}$ for $Z_q \cdot Z_m$.
Hence, we choose $|\tilde{p}| = 2 \,\text{GeV}$ as our optimal
matching scale.
Our preliminary results are
\begin{equation}
Z_q'(\tilde{p} = 2\,\text{GeV}) = 0.9810(46) \,,
\qquad
Z_q \cdot Z_m (\tilde{p} = 2\,\text{GeV}) = 1.0551(52) \,.
\end{equation}
We plan to cross-check these results against those obtained using the
bilinear operators in near future.

\section{Acknowledgments}
W.~Lee is supported by the Creative Research
Initiatives program (2012-0000241) of the NRF grant funded by the
Korean government (MEST).
W.~Lee acknowledges support from the KISTI supercomputing
center through the strategic support program [No. KSC-2011-G2-06].
Computations were carried out in part on QCDOC computing facilities of
the USQCD Collaboration at Brookhaven National Lab, on GPU computing
facilities at Jefferson Lab, on the DAVID GPU clusters at Seoul
National University, and on the KISTI supercomputers. The USQCD
Collaboration are funded by the Office of Science of the
U.S. Department of Energy.

\end{document}